\documentclass[aps,prb,twocolumn,superscriptaddress,showpacs,footnotebib]{revtex4-1}
\usepackage{amsmath}
\usepackage{color}
\usepackage{graphicx}
\usepackage{epstopdf}
\usepackage{epsfig}

\usepackage{amsfonts}
\usepackage[naturalnames]{hyperref}
\usepackage{hypcap}
\usepackage{verbatim}
\usepackage{tabularx}
\usepackage{bbm}
\usepackage{esvect}
\usepackage{subfigure}

\begin{document}
\graphicspath{{newfig/}}
\title{Three-dimensional time reversal invariant topological superconductivity in doped chiral topological semimetals}
\author{Yingyi Huang}
\affiliation{Institute for Advanced Study, Tsinghua University, Beijing 100084, China}
\author{Shao-Kai Jian}
\affiliation{Institute for Advanced Study, Tsinghua University, Beijing 100084, China}
\affiliation{Condensed Matter Theory Center Maryland, Department of Physics, University of Maryland, College Park, MD 20742, USA}
\date{\today}
\begin{abstract}
Chiral topological semimetals host multifold degenerate band crossing points under the protection of crystalline symmetries.  
In this paper, we suggest that the recently discovered chiral topological semimetals in space group 198, parts of which are superconducting on doping, can be new candidates of time reversal invariant topological superconductors. 
By investigating the Fermi surfaces around the band crossing points that carry nonzero Chern numbers, we clarify how the nontrivial topology of chiral topological semimetals affects their superconducting state and show the existence of topological superconductivity in $s_\pm$-wave pairing with surface Majorana fermions. 
We further demonstrate that the topological superconductivity is favored by the inter-unit-cell phonon-mediated electron-electron interaction.
\end{abstract}
\maketitle

\section{Introduction}
 
Over the past decade, the interplay of superconducting pairing and band topology has paved a route to topological superconductors (TSCs) hosting Majorana modes, which have potential applications in topological quantum computation~\cite{Nayak2008Non-Abelian, DasSarma2015Majorana,Alicea2012New,Elliott2015Colloquium,Stanescu2013Majorana,Leijnse2012Introduction,Beenakker2013Search,Lutchyn2017Realizing,jiang2013non,sato2016majorana,Sato2016Topological,Aguado2017Majorana}. 
For instance, this intriguing interplay can be realized by the proximity of conventional $s$-wave superconductor (SC) and the one-dimensional (1D) spin-orbit coupled semiconductor~\cite{Sau_semiconductor_heterostructures,Roman_SC_semi,Gil_Majorana_wire} or 2D  topological insulator~\cite{FuKane_SC_STI}. 
Despite the fact that experimental progress have been made~\cite{Mourik2012Signatures,Zhang2017Quantized,Nichele2017Scaling}, the heterostructures make  experimental confirmation and further application complicated. 
On the other hand, superconductivity is found to exist in topological (crystalline) insulator under doping, pressure, and/or low temperature. 
Famous examples include Cu$_x$Bi$_2$Se$_3$~\cite{FuBerg2010,PhysRevLett.107.217001,PhysRevLett.110.117001} and Sn$_{1-x}$In$_x$Te~\cite{novak2013unusual},
%This mechanism also applies to 
as well as gapless topological matters (e.g., doped Weyl and Dirac semimetals)~\cite{hosur2014time,PhysRevLett.115.187001}. 
Although zero-bias conductance peaks have been experimentally observed in iron-based superconductors~\cite{Wang333, Zhang182,Chen_2019,2018arXiv181208995M,PhysRevX.8.041056}, further investigation on superconducting topological matter is necessary.

\begin{figure}[t]
\centering
\subfigure[(a)]{\includegraphics[width=0.12\textwidth]{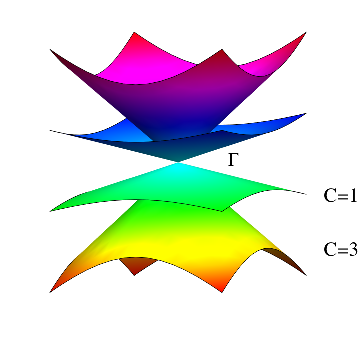}}
\subfigure[(b)]{\includegraphics[width=0.14\textwidth]{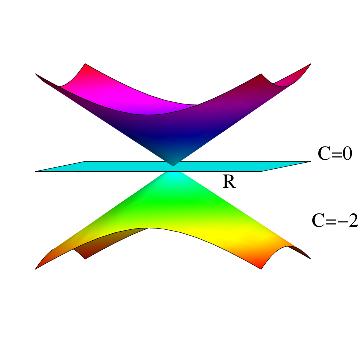}} \quad
\subfigure[(c)]{\includegraphics[width=0.16\textwidth]{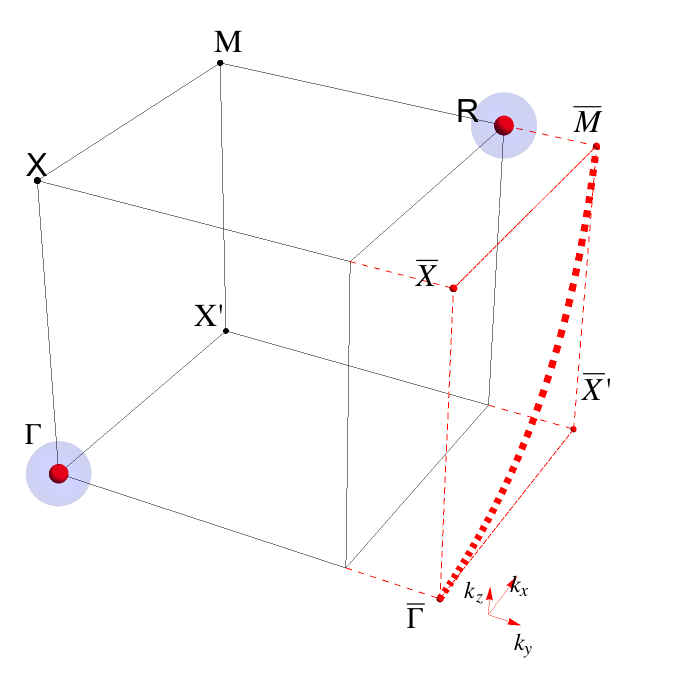}}
\caption{\label{Fig1} (Color online)Energy dispersion for the chiral topological semimetal in SG 198 with (a) fourfold degeneracy at the $\Gamma$ point and (b) sixfold degeneracy, i.e., two threefold degeneracy at the $R$ point (the figure shows one of them). (c) The BZ with Fermi surfaces around $\Gamma$ and $R$ point paired by $s_\pm$-wave pairing SC. The projected BZ of the (001) surface is marked in red lines with the projected Fermi arcs in the dashed line. 
}
\end{figure}

Recently, another type of gapless topological matters with nonsymmorphic space group (SG) symmetry, dubbed as ``chiral topological semimetals" or chiral crystals~\cite{bradlyn2016beyond,chang2017unconventional, tang2017multiple}, has attracted a lot of experimental attention~\cite{rao2019observation,takane2019observation,sanchez2019topological,schroter2019chiral, lv2019observation}. 
The chiral topological semimetals can be viewed as natural generalizations of the well-studied Weyl semimetals. 
In particular, the spin-$1/2$ vector $\bf{S}$, in the low-energy Hamiltonian of Weyl semimetals $H=\delta \bf{k}\cdot \bf{S}$, is replaced by spin-1 or -3/2 matrices in chiral topological semimetals~\cite{bradlyn2016beyond,chang2017unconventional, tang2017multiple}. 
However, the chiral topological semimetals have distinguished bulk and edge topological properties. Different from the chiral fermions in Weyl semimetals with a topological charge (Chern number) $\mathcal C=\pm1$, the chiral fermions in chiral topological semimetals carry topological charge $\mathcal C$ larger than 1. 
For instance, the chiral topological semimetals in SG 198 host fourfold and sixfold degenerate chiral fermions, as shown in Figs.~\ref{Fig1}(a) and (b). 
While the band crossings of the Weyl semimetals are in general located at positions away from the high-symmetry points of the Brillouin zone (BZ),  the multifold degenerate band crossing points of the chiral topological semimetals are located at high-symmetry points.

In analog with TSC in doped Weyl semimetals, it is natural to expect interesting superconducting states to emerge in these chiral topological semimetals upon doping.  
In this paper, we address the effect of the nontrivial topology, i.e., Fermi surfaces with nonzero Chern numbers due to the multifold degenerate points, on the superconducting properties. 
That the multifold degenerate crossings are located at high symmetric points (e.g., in SG 198, $\mathcal C=\pm 4$ at $\Gamma$ and $R$ points) is a unique property of the chiral topological semimetals distinguished from the Weyl semimetals and leads to the possibility that an $s$-wave pairing can be nontrivial (more precisely, $s_\pm$-wave pairing). 
Also, the high Chern number in normal state can give a large topological number in the SC state of chiral topological semimetals, which is distinguished from that of Weyl semimetals. 

Recently, superconducting states of multifold linear band crossing point in the chiral topological semimetals have been theoretically investigated~\cite{boettcher2020interplay, link2020d,sim2019triplet,lin2019chiral}. 
However, the analysis~\cite{boettcher2020interplay, link2020d} either focused on unconventional superconductivity arising from the multifold band crossing point near $\Gamma$ point, or focused on the three band crossings~\cite{sim2019triplet,lin2019chiral} at $R$ point and did not consider both topological nontrivial Fermi surfaces surrounding the $\Gamma$ and $R$ points simultaneously. Given the distinct topological properties that appear at the $\Gamma$ and $R$ points, and the accumulating experimental evidence of superconducting states in these chiral topological semimetals, it is of great importance and interest to ask whether nontrivial topological states can emerge.
 
In this paper, we answer this question in the affirmative. We propose that the doped chiral topological semimetals of SG 198 are promising candidates for realizing the time-reversal invariant TSC. On slight doping, the Fermi surfaces with nontrivial Chern number appear around both $\Gamma$ and $R$ points in the BZ of the chiral topological semimetals. We find that a momentum-dependent superconducting pairing, whose pairing amplitude changes sign between the $\Gamma$ point and the $R$ point, can realize the TSC.  By carefully examining a tight-binding model consistent with the underlying nonsymmorphic SG symmetry, we show that an $s_\pm$-wave superconducting pairing is fully gapped and topologically nontrivial with a large winding number $\nu = 4$, and we also show explicitly the resultant Majorana surface states. We finally discuss the effect of interactions in the material that favors the topological superconducting state.

\section{Topologically nontrivial Fermi surfaces in chiral semimetals}
 
We begin with an analysis of the topology of multiple band crossing points. The non-symmorphic SG symmetry in SG 198 leads to multiple band crossing points at two time reversal invariant points: the BZ center $\Gamma$ point and the BZ corner $R$ point. The existence of the two multifold degenerate nodes renders nonzero Chern numbers of the bands below (or above) the Fermi level of these points.  The topological charge of the bands below the Fermi level of the nodes is $3$ and $1$ at $\Gamma$ point and $-2$ and $-2$ at $R$ point, as shown in Figs.~\ref{Fig1}(a) and \ref{Fig1}(b). The total topological charge at $\Gamma$ and $R$ is $\pm4$, respectively, and their sum is zero, which is consistent with the Nielsen-Ninomiya no-go theorem~\cite{nielsen1981no}.

With a finite carrier density, disconnected Fermi surfaces surrounding the $\Gamma$ and $R$ points appear in the chiral topological semimetal as shown in Fig.~\ref{Fig1}(c). 
We consider a case in which the chemical potential locates above the degenerate nodes at both $\Gamma$ and $R$ points so that there are four Fermi surfaces with nonzero Chern number: two enclose the $\Gamma$ point with Chern number  $\mathcal{C}=3$ and $\mathcal{C}=1$, and the other two enclose the $R$ point with $\mathcal{C}=-2$ and its own time-reversal partner.

According to topological classification~\cite{chiu_RMP_16}, a 3D time-reversal invariant superconductor in class DIII is characterized by an integer ($\mathbb{Z}$) topological invariant, which can be written as a winding number over the entire momentum space. The winding number for a 3D TSC is defined as~\cite{schnyder2008classification}
\begin{eqnarray}
\nu=\frac1{24\pi^2}\int d^3{\bf k}\epsilon^{ijk}{\rm Tr}\left[{Q^\dagger_{\bf k}\partial_iQ_{\bf k}Q^\dagger_{\bf
k}\partial_jQ_{\bf k}Q^\dagger_{\bf k}\partial_kQ_{\bf k}}\right].
\label{eq:Windingnumber}
\end{eqnarray}
Here,  $Q_{\bf k}$ is related to the projection operator of occupied state $P_{\bf k}=\sum\limits_{n\in filled}|u_{n\bf k}\rangle\langle u_{n\bf k}|$ as $Q_{\bf k}=1-2P_{\bf k}$ with $|u_{n\bf k}\rangle$ being the $n$-th Bloch wave function at ${\bf k}$. 

To relate the winding number to the Chern number of the Fermi surface, we consider a simple case: When the superconducting gap is much smaller than the Fermi energy, the topological properties of the superconductor pairing are completely determined by the Fermi surfaces in the normal state~\cite{QiHughesZhang10}.  
Under this weak pairing assumption, the winding number for a TSC reduces to
\begin{equation}
\label{eq:topoin}
\nu=\frac{1}{2}\sum\limits_{j\in \text{FS}} \mathcal C_{j}\text{sgn}(\Delta_j),
\end{equation}
where $\mathcal C_j$ is the Chern number carried by the $j$-th Fermi surface and $\Delta_j$ is the SC pairing of the $j$-th Fermi surface.
Due to the time-reversal symmetry, the SC pairing gap function $\Delta_j$ on the $j$-th surface is real and is well defined for a gapped system.

Apparently, it is easy to show that the conventional $s$-wave pairing is topological trivial. 
The $s$-wave SC pairing gap functions on all the Fermi surfaces are all positive (or equivalently all negative). 
Since the total Chern number for all the Fermi surfaces in the BZ is zero guaranteed by the no-go theorem~\cite{nielsen1981no}, the winding number is zero for a uniform SC pairing potential.
As suggested by Eq.~(\ref{eq:topoin}), a possible way to obtain a nonzero winding number is that the Fermi surfaces with opposite Chern number form pairing functions with opposite sign. Since in the chiral topological semimetal the Fermi surfaces near $\Gamma$ and $R$ points, two far apart points in the BZ, have opposite Chern number as shown in Fig.~\ref{Fig1}, if the SC pairings of these two sets of Fermi surface feature an opposite sign, one can immediately get a nontrivial winding number. 
Therefore, the $s_\pm$-wave SC pairing, which does not break crystalline rotational symmetry but changes sign across the BZ from $\Gamma$ point to $R$ point, naturally leads to a TSC.
More precisely, in the $s_\pm$-wave pairing, the electrons from each Fermi surface form spin-singlet pairs (i.e., Kramers pairs); in particular, in the chiral semimetal there are four Fermi surfaces forming four pairing functions $\Delta_j$, and the signs of the pairing functions are opposite for those enclosing the $\Gamma$ and $R$ points. Then, according to Eq.~(\ref{eq:topoin}), the winding number is $\nu=4$. 

The singlet pairing function can be written in a form as
\begin{equation} 
\Delta(\textbf{k})=i d(\textbf{k})\sigma_y. 
\label{eq:SC}
\end{equation}
where $\sigma$ denotes spin, and $d(\textbf{k})$ is an even functions of $\textbf{k}$ and forms a irreducible representation of the point group $\mathbbm{T}$.  
For $s_\pm$ wave SC pairing, the form factor can be chosen to be $d(\textbf{k})= \Delta_0\left(\cos(k_x)+\cos(k_y)+\cos(k_z)\right),$ which changes sign from $\Gamma$ to $R$ point. We show in the next section that, as anticipated, the $s_\pm$-wave SC pairing gives a TSC with $\nu=4$.

\section{s$_\pm$ superconducting pairing}
\begin{figure}
\centering
\subfigure[(a)]{\includegraphics[width=0.20\textwidth]{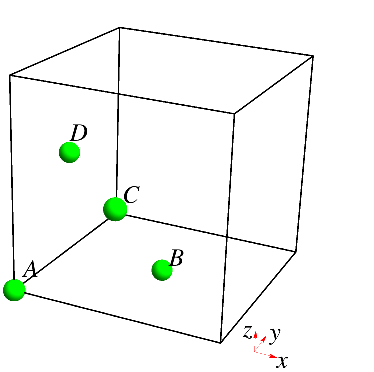}}
\subfigure[(b)]{\includegraphics[width=0.24\textwidth]{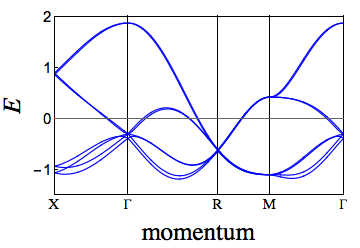}}
\caption{ (Color online) (a) A schematic plot of the lattice structure in real space within a unit cell. The orbitals (sublattices) $A$(0,0,0), $B(\frac{1}{2}, \frac{1}{2}, 0)$, $C(\frac{1}{2}, 0, \frac{1}{2})$, and $D(0, \frac{1}{2}, \frac{1}{2})$ in the same unit cell are marked in green. 
(b) The band dispersion of RhSi~\cite{chang2017unconventional} along a typical high symmetric cut in the BZ.}
\label{Fig2}
\end{figure}

For definiteness, we study a tight-binding Hamiltonian respecting the time-reversal symmetry $\mathcal{T}$ and the nonsymmorphic symmetries of SG 198. 
The insulating filling constraint for spin-orbit coupled insulators in SG 198 is $ 8\mathbb{Z}$ according to Ref.~\onlinecite{watanabe2015filling}. 
It indicates that the tight-binding model should be described by an $8\times 8$ matrix. 
The Hamiltonian can be written in a basis of spin $\sigma$ and four Wannier orbitals in a unit cell shown in Fig.~\ref{Fig2}(a).  
The tight-binding model $H_0$ was derived in Ref.~\cite{chang2017unconventional}, and the full expression is long and tedious, so we give the main result here and leave details to the Supplemental Material~\cite{suppl1}.
We want to emphasize that the non-symmorphic symmetries are critical in determining the tight-binding model in which the dispersions around the $\Gamma$ and $R$ points need to be correctly accounted for to realize the nontrivial topological state.

Using the lattice model $H_0$, we reproduce the band dispersion of RhSi in Fig.~\ref{Fig2}(b). 
We can see that there is a sixfold degeneracy at $\Gamma$ and a fourfold degeneracy at $R$, in consistence with the angle-resolved photoemission spectroscopy (ARPES) measurement~\cite{rao2019observation, takane2019observation, sanchez2019topological}.  
In addition, through Eq.~\eqref{eq:Windingnumber}, we calculate the winding number of Hamiltonian $H_0$ in the presence of the $s_\pm$ pairing function, i.e., the winding number of the BdG Hamiltonian
\begin{equation}
	\label{eq:HBdG}
	\textit{H}_\text{BdG}(\bf{k})=\begin{pmatrix} H_0(\textbf{k})-\mu & \Delta(\textbf{k})\\ \Delta^\dag(\textbf{k}) & -H_0^T(-\textbf{k})+\mu\end{pmatrix},
\end{equation}
which leads to %the calculation of Eq.~\ref{eq:Windingnumber} which results 
$\nu=4$, in agreement with the analytical analysis.  
Note that the results only depend on the Fermi surface topology, but not on the fine-tuning of parameters.

\begin{figure}[htbp] \centering 
\subfigure[(a)]{\includegraphics[width=0.20\textwidth]{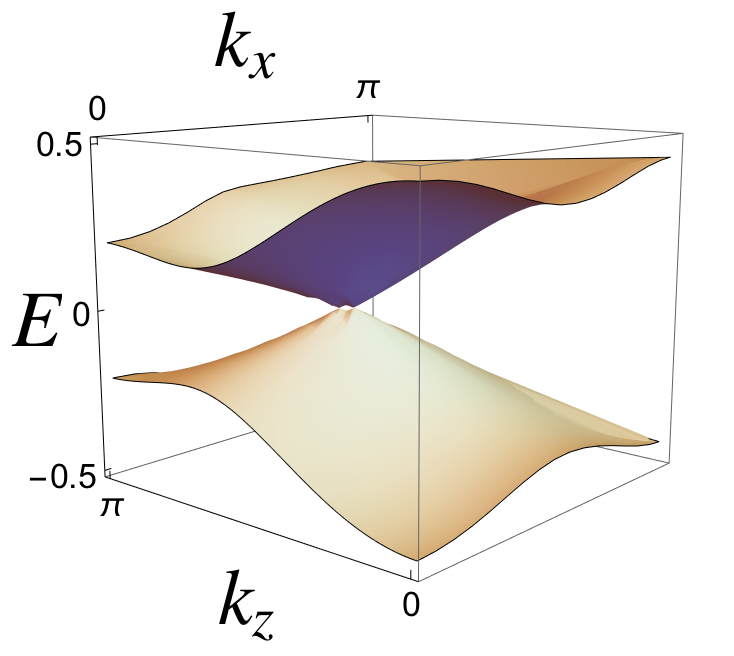}}
\subfigure[(b)]{\includegraphics[width=0.24\textwidth]{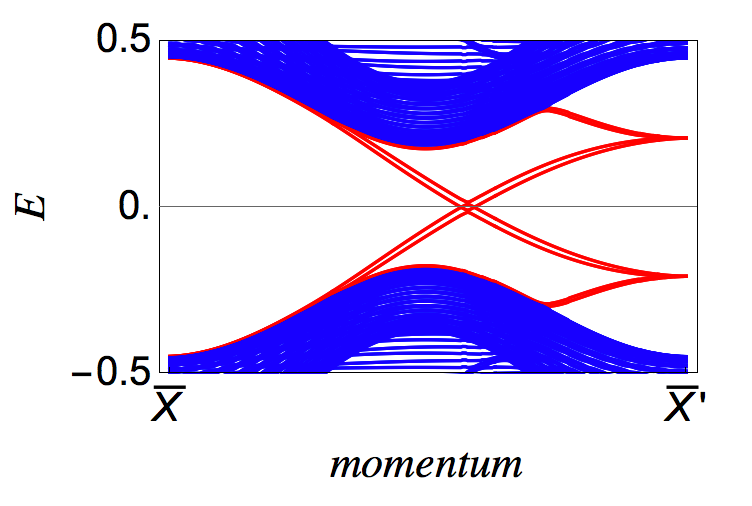}}
\caption{ (Color online) (a) The surface states on $k_x$-$k_z$ surface BZ and (b) the BdG spectra along an arc connecting $\bar{X}$ and $\bar{X}'$ with SC pairing amplitude $\Delta=0.2$. The solid red lines indicate the surface Majorana cone. The parameters are the same as in Fig.~\ref{Fig2}. } 
\label{Fig3}
\end{figure}

In the presence of the $s_\pm$-wave SC pairing, we found that the chiral topological semimetal becomes a fully gapped TSC by numerically calculating the energy spectrum the BdG Hamiltonian [Eq.~\eqref{eq:HBdG}]. 
The critical doping (chemical potential) is discussed in the Supplemental Material~\cite{suppl3}. 
Due to the bulk-edge correspondence, we expect the existence of gapless modes at a 2D surface as a manifest of the nontrivial topological property of a 3D TSC. To show this, we consider a geometry with periodic boundary conditions in the $x$ and $z$ directions and an open one in the $y$ direction. As shown in Fig.~\ref{Fig3}(a), there are surface states which have well-defined $k_x$ and $k_z$ momenta. 
One can see that the energies of such surface states become zero at some $k_x$ and $k_z$ momenta, forming two Majorana cones. 
In the full surface BZ, there are eight Majorana cones in total, among which four come from the surface terminated at $+y$ and the other four come from the surface terminated at $-y$.
We further plot the band spectrum along the minimal of the surface states at different $k_z$'s on the $k_x$-$k_z$ plane. 
Figure~\ref{Fig3}(b) shows that there are two pairs of helical Majorana states crossing the bulk superconducting gap consisting with two Majorana cones shown in Fig.~\ref{Fig3}(a).  
Note that the momentum in Fig.~\ref{Fig3}(b) is not a straight line connecting $\bar X$ and $\bar X'$ but an arc determined by the momentum of the minimal energy state among all $k_z$'s for every fixed $k_x$ connecting $\bar X$ and $\bar X'$.

\section{Interaction effect} 
As we have shown that a doped chiral topological semimetal with an $s_\pm$-wave pairing can be a promising platform for realization a TSC, it is necessary to examine the stability of the pairing state in the presence of interactions.
We will investigate the microscopic interaction that favors the $s_\pm$-wave state using the fluctuation-exchange approach~\cite{hosur2014time} and consider on-site interaction and momentum-dependent interaction up to first order. 
Three kinds of interactions will be included: The effective intra-unit-cell intra-orbital interaction $U$, the effective intra-unit-cell inter-orbital interaction $V$, and effective intra-orbital interaction between adjacent unit cells $W$. 
In general, the effective interaction parameters are given by the difference between electron-phonon coupling $I_\text{ph}$ and Coulomb repulsion $I_\text{Cl}$ as $I=I_\text{ph}-I_\text{Cl}$, $I=U, V, W$. The interaction is written as
\begin{equation}
\begin{aligned}
\label{eq:int}
H_\text{int}=&-U\sum\limits_{i, l} n_{il} n_{il}-V\sum\limits_{i, l\neq l'} n_{il} n_{il'} \\
&-W\sum\limits_{\langle i, i'\rangle, l} n_{il} n_{i'l},
\end{aligned}
\end{equation}
where $l$ indices are summed over all the orbitals in one unit cell ($l=A, B, C, D$) as shown in Fig.~\ref{Fig2}(a); $n_{il}=\sum\limits_{\sigma}c^\dag_{il\sigma}c_{il\sigma}$ is the electron density with spin $\sigma=\uparrow,\downarrow$ locating at sublattice $l$ in the unit cell $i$.

In the BCS theory, if we start from a free electron gas and turn on an attractive interaction, the system becomes unstable and the electrons group themselves in pairs. 
The relevant interaction is the Cooper channel, where two electrons $\psi_\textbf{k}$ and $\psi_\textbf{-k}$ scatter toward final states $\psi_{\textbf{k}'}$ and $\psi_{\textbf{-k}'}$. 
Thus, the inter-unit-cell interaction $W$ is necessary for the $s_\pm$-wave pairing because only the scattering process that connects the $\Gamma$ and $R$ points at different momenta $\textbf{k}=(0,0,0)$ and $\textbf{k}'=(\pi, \pi, \pi)$ can induce a momentum-dependent pairing.

Since the chiral topological semimetal preserves the time-reversal symmetry, a natural route to realize it is through pairing between electrons with the momenta at the Fermi surfaces that are Kramers partners as we have discussed before. 
The interaction projected onto the Fermi surfaces takes the following form:
\begin{equation}
\chi^{ij}(\textbf{k},\textbf{k}')=\langle \psi_{\textbf{k}',j}|\otimes \langle \mathcal{T}\psi_{\textbf{k}',j}|H_{int}(\textbf{k},\textbf{k}') |\mathcal{T}\psi_{\textbf{k},i}\rangle\otimes|\psi_{\textbf{k},i}\rangle,
\end{equation}
where $|\psi_{\textbf{k},i}\rangle$ being one of the following four wave functions at the Fermi surfaces $\{|\psi_{\Gamma,3/2}\rangle, |\psi_{\Gamma,1/2}\rangle, |\psi_{R,-2}\rangle, |\mathcal{T}{\psi_{R,-2}}\rangle^*\}$. The detailed construction of these wave functions is presented in the Supplemental Material~\cite{suppl2}.

With the effective interactions, we obtain the pairing order parameter by solving the linearized gap equation
\begin{equation}
\sum\limits_j\int_k'\delta(v_j \delta k')\chi^{ij}(k,k')\Delta_j(k') =\lambda\Delta_i(k),
\end{equation}
where $\int_k'\equiv\int\frac{d^3 k'}{(2\pi)^3}$ is the sum over the Fermi surfaces. 
The most negative eigenvalue $\lambda$ gives the highest transition temperature $T_c\propto \exp(-1/\lambda )$ and determines the dominant pairing. 
The corresponding eigenvector gives the pairing function $\Delta_j$, in particular, the signs of the pairing function that combining the information of nonzero Chern number of Fermi surfaces determined the winding number through Eq.~\eqref{eq:topoin}.

\begin{figure}[htbp] \centering 
\subfigure[(a)]{\includegraphics[width=0.22\textwidth]{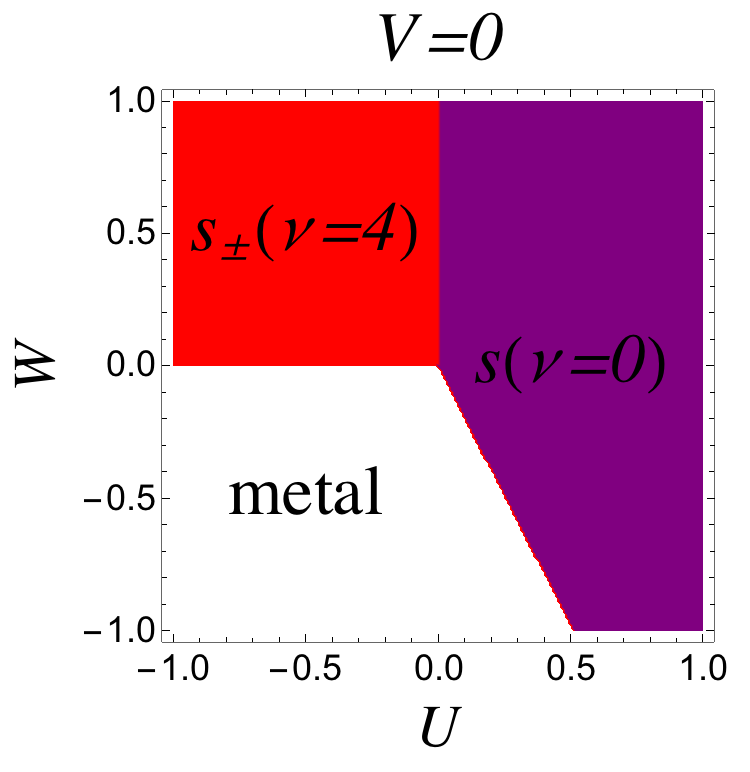}}
\subfigure[(b)]{\includegraphics[width=0.22\textwidth]{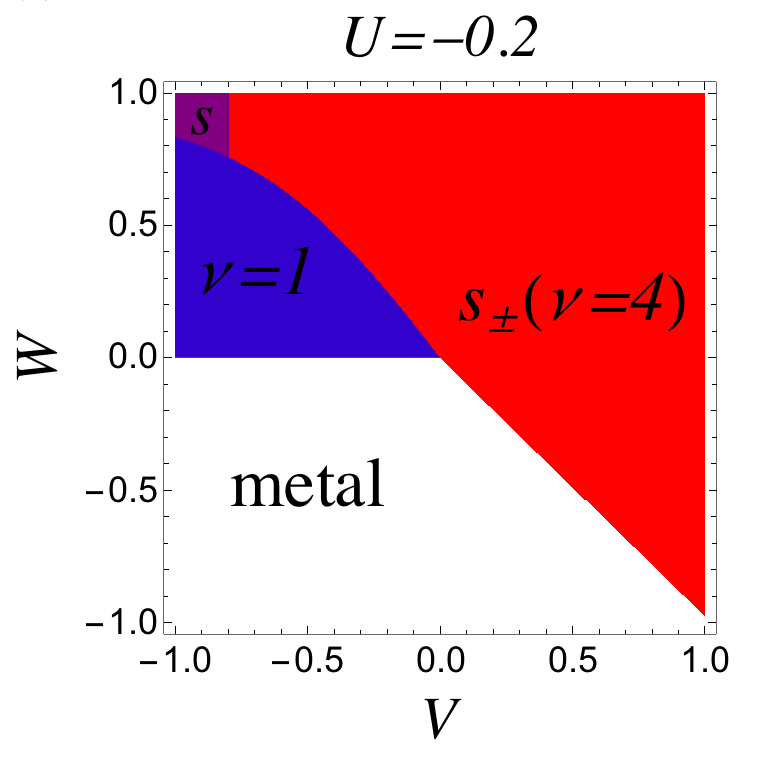}}
\caption{\label{Fig4}(Color online) The phase diagram for inter-unit cell interaction $W$ and on-site (a) intra-orbital interaction $U$ and (b) inter-orbital interaction $V$ respectively. Only the most leading channel at each point is indicated: $s_\pm$-wave pairing(red), $s$-wave pairing(purple), and  $\nu=1$ phase(blue).}
\end{figure}

The phase diagrams as a function of interactions are shown in Fig.~\ref{Fig4}. First, we consider the case with $V=0$ in Fig.~\ref{Fig4}(a).  We find the topological nontrivial $s_\pm$-wave pairing phase at $W>0$ and $U<0$. In this region, the $s_\pm$-wave pairing is favored over the conventional $s$-wave pairing due to the on-site repulsion from $U$. The inter-unit-cell interaction $W$ is a result of competition between the electron-electron repulsive Coulomb interaction and the phonon-mediated attractive interaction.
Since Coulomb interaction decreases inversely with distance, whereas the phonon mode connecting two unit cell can directly contribute to attractive interaction, one might expect $W>0$ in some of the materials.
The conventional $s$-wave SC phase requires larger phonon-mediated electron-electron attraction than $s_\pm$-wave SC phase. 
So, we consider the plot of $V$ and $W$ with repulsive $U=-0.2$ in Fig.~\ref{Fig4}(b). The region for $s$-wave SC state diminishes in the presence of $U<0$ and $V<0$, which is replaced by $s_\pm$-wave pairing. 
This suggests that to realize the topological phase, the repulsive on-site interaction is useful to suppress the trivial $s$-wave pairing.
Moreover, a new nontrivial phase with $\nu=1$ emerges as a result of the SC sign changing between $\mathcal C=1$ and $\mathcal C=3$ phases at the $\Gamma$ point. 
Further understanding of this new phase is needed.

\section{Conclusion}
We have revealed that 3D time reversal invariant TSC phase can be realized in a doped chiral topological semimetal. 
In particular, we have shown the existence of TSC phase with $\nu=4$ in chiral topological semimetals of SG 198 by calculating the topological invariant and the BdG spectrum. 
We present a (incomplete) survey of possible candidates to realize our proposals.

The chiral topological semimetals can be realized in materials with time-reversal symmetry and spin-orbit couplings, and their host of crystalline symmetry-protected fermionic excitations are stabilized by nonsymmorphic symmetries~\cite{bradlyn2016beyond}.
The most considered chiral crystals are in SG 198 primitive cubic lattice~\cite{chang2017unconventional, tang2017multiple}. 
Recently, chiral topological semimetals have been found with the observation of long Fermi arcs by ARPES in several materials of SG 198, including CoSi~\cite{rao2019observation,takane2019observation,sanchez2019topological}, RhSi~\cite{ sanchez2019topological}, PtGa~\cite{yao2020observation}, PdGa~\cite{schroter2020observation}, AIPt~\cite{schroter2019chiral}, and PdBiSe~\cite{lv2019observation}. 
More importantly, superconducting phase transitions were reported in several compounds in the cubic SG 198. 
AuBe, a transition metal compound with multifold degenerate nodes in calculated electronic structure~\cite{rebar2019fermi}, is reported to be a noncentrosymmetric superconductor subjecting to applied magnetic field from muon spin relaxation measurement~\cite{amon2018noncentrosymmetric,singh2019type,beare2019mu}. 
In an Ullmannite-type compound, PdBiSe, ARPES measurement has observed multiple unconventional fermions~\cite{lv2019observation} and electrical transport and AC susceptibility measurements have found a superconducting transition around 1.8 K~\cite{2015JPhCS.592a2069J}. 

Our work on realizing TSC in topological chiral semimetal, therefore, suggest studies and experimental investigations of TSCs in a large family of materials.
The possible Majorana excitations hosted in the chiral topological semimetal can be detected by ARPES or the scanning tunneling microscope.

\begin{acknowledgments}
We thank Zhong Wang for suggesting this project to us. Y. H. is indebted to Baiqing Lv, Lingyuan Kong, Wen Huang, Zhongbo Yan, Ai-Lei He, and Ching-Kai Chiu for helpful discussions. 
 \end{acknowledgments}
\bibliography{weyl}
\onecolumngrid
\vspace{1cm}
\begin{center}
{\bf\large Supplemental Material}
\end{center}
\vspace{0.5cm}

\setcounter{section}{0}
\setcounter{secnumdepth}{3}
\setcounter{equation}{0}
\setcounter{figure}{0}
\renewcommand{\theequation}{S-\arabic{equation}}
\renewcommand{\thefigure}{S\arabic{figure}}
\renewcommand\figurename{Supplementary Figure}
\renewcommand\tablename{Supplementary Table}
\newcommand\Scite[1]{[S\citealp{#1}]}
\makeatletter \renewcommand\@biblabel[1]{[S#1]} \makeatother
%%%%%%%%%%%%%%%%%%%%%%%%%%%%%%%%%%
% The supplementary text starts here
%%%%%%%%%%%%%%%%%%%%%%%%%%%%%%%%%% 

\section{The effective tight-binding model}
To characterize the chiral topological semimetals protecting by non-symmorphic space group symmetries, it is necessary to construct a effective tight-binding model instead of $\bf{k}\cdot\bf{p}$ models around high-symmetry points. Although the $\bf{k}\cdot\bf{p}$ models manifest the degeneracies of the high-spin fermions hosting in a material, it does not determine all the key properties of the material. The band dispersion away from the high-symmetry point, and any band crossings in the dispersion are also fundamental properties of the material. 
 
While the $\bf{k}\cdot\bf{p}$ Hamiltonian is constructed from the irreducible representations of the little group at high-symmetry invariant, the tight-binding model is based on the representations of the space group. In particular,  the materials in SG 198 are protected by a diagonal cubic threefold rotation $C_{3,111}$ and two two-fold screw rotation symmetries $C_{2x}$ and $C_{2z}$. Their generators are described by point group $\mathbbm{T}$ (No. 23) in combination with the translational parts.  In the presence of the translational parts, the generators at different high-symmetric points are distinguished. In particular, the generators at $\Gamma$ point are
$s_{2x}=\left\{C_{2x}|\frac{1}{2}\frac{1}{2}0\right\},s_{2z}=\left\{C_{2z}|\frac{1}{2}0\frac{1}{2}\right\}$, and $s_{3,111}=\left\{C_{3,111}|000\right\}$ with $\mathcal{K}$ being the complex conjugation. At $R$ point, the generators become $s_{2x}=\left\{C_{2x}|\frac{1}{2}\frac{3}{2}0\right\},s_{2z}=\left\{C_{2z}|\frac{3}{2}0\frac{1}{2}\right\}$, and $s_{3,111}=\left\{C_{3,111}|010\right\}$. 
 We can write down the T representations of point group 23 :
\begin{align}
&\{\sigma_x,\sigma_y,\sigma_z\}\\ \nonumber
&\{\tau_x,\tau_x\mu_x,\mu_x\}\\ \nonumber
&\{\tau_y,\tau_x\mu_y,\tau_z\mu_y\}\\ \nonumber
&\{\tau_y\mu_z,\tau_y\mu_x,\mu_y\}\\ \nonumber
&\{\tau_x\mu_z,-\tau_y\mu_y,\tau_z\mu_x\}.
\end{align}
Here, the Pauli matrix $\sigma$ represents spin and $\tau$ ($\mu$) describes the four Wannier orbitals in a unit-cell.  To be specific, $\tau^x$ represents hopping between orbitals $A$ and $B$, or between orbitals $C$ and $D$. Similarly, $\mu^x$ represents hopping between $A$ and $C$, or $B$ and $D$.  Under the basis $(c^A, c^C, c^B, c^D)$, the operations can be written as $s_{2x}=i\tau^x\sigma^x\otimes(k_{y,z}\rightarrow-k_{y,z}), s_{2z}=i\mu^x\sigma^z\otimes(k_{x,y}\rightarrow-k_{x,y}),  \mathcal{T}=i\sigma^y\mathcal{K}\otimes(k_{x,y,z}\rightarrow-k_{x,y,z})$
and
$C_{3,111}=\left(\begin{array}{cccc}1 & 0 & 0 & 0 \\0 & 0 & 1 & 0 \\0 & 0 & 0 & 1 \\0 & 1 & 0 & 0\end{array}\right)e^{ \pi i\sigma_{111}/3}\otimes(k_x\rightarrow k_y, k_y\rightarrow k_z, k_z\rightarrow k_x)$ with
 $\sigma_{111}=(\sigma_x+\sigma_y+\sigma_z)/\sqrt{3}$.

Based on the irreducible representation theory, we can obtain the terms preserving space group symmetry as product of T representation of point group and momentum. With momentum up to second order, the tight-binding model firstly presented in Ref.~\onlinecite{chang2017unconventional} is 
%\begin{equation}
\begin{align}
\textit{H}_0&=v_1\left(\tau^x\cos\frac{k_x}{2}\cos\frac{k_y}{2}+\tau^x\mu^x\cos\frac{k_y}{2}\cos\frac{k_z}{2}+\mu^x\cos\frac{k_z}{2}\cos\frac{k_x}{2}\right)\\ \nonumber
&+v_p\left(\tau^y\mu^z\cos\frac{k_x}{2}\sin\frac{k_y}{2}+\tau^y\mu^x\cos\frac{k_y}{2}\sin\frac{k_z}{2}+\mu^y\cos\frac{k_z}{2}\sin\frac{k_x}{2}\right)\\ \nonumber
&+v_{r1}\left(\tau^y\mu^z\sigma^y\cos\frac{k_x}{2}\cos\frac{k_y}{2}+\tau^y\mu^x\sigma^z\cos\frac{k_y}{2}\cos\frac{k_z}{2}+\mu^y\sigma^x\cos\frac{k_z}{2}\cos\frac{k_x}{2}\right)\\ \nonumber
&+v_{r2}\left(\tau^y\sigma^z\cos\frac{k_x}{2}\cos\frac{k_y}{2}+\tau^x\mu^y\sigma^x\cos\frac{k_y}{2}\cos\frac{k_z}{2}+\tau^z\mu^y\sigma^y\cos\frac{k_z}{2}\cos\frac{k_x}{2}\right)\\ \nonumber
&+v_{r3}\left(\tau^y\mu^z\sigma^x\sin\frac{k_x}{2}\sin\frac{k_y}{2}+\tau^y\mu^x\sigma^y\sin\frac{k_y}{2}\sin\frac{k_z}{2}+\mu^y\sigma^z\sin\frac{k_z}{2}\sin\frac{k_x}{2}\right)\\ \nonumber
&+v_{s1}\left(\tau^x\sigma^x\sin\frac{k_x}{2}\cos\frac{k_y}{2}+\tau^x\mu^x\sigma^y\sin\frac{k_y}{2}\cos\frac{k_z}{2}+\mu^x\sigma^z\sin\frac{k_z}{2}\cos\frac{k_x}{2}\right)\\ \nonumber
&+v_{s2}\left(\tau^x\sigma^y\cos\frac{k_x}{2}\sin\frac{k_y}{2}+\tau^x\mu^x\sigma^z\cos\frac{k_y}{2}\sin\frac{k_z}{2}+\mu^x\sigma^x\cos\frac{k_z}{2}\sin\frac{k_x}{2}\right)\\ \nonumber
&+v_{s3}\left(\tau^x\mu^z\sigma^z\cos\frac{k_x}{2}\sin\frac{k_y}{2}-\tau^y\mu^y\sigma^x\cos\frac{k_y}{2}\sin\frac{k_z}{2}+\tau^z\mu^x\sigma^y\cos\frac{k_z}{2}\sin\frac{k_x}{2}\right)\\ \nonumber
&+v_2[\cos(k_x)+\cos(k_y)+\cos(k_z) ].\nonumber
 %\end{split}
 \end{align}
 %\end{equation}
The parameters in Figs.~\ref{Fig2} and \ref{Fig3} are chosen following Ref.~\onlinecite{chang2017unconventional} as: $v_1=0.55,v_p=-0.76,v_{s1}=-0.04, v_{s2}=0,v_{s3}=0,v_{r1}=0,v_{r2}=-0.03,v_{r3}=0.01,v_2=0.16,\mu=0.2$.

\section{The critical doping}
 In undoped samples, the density of states in such semimetallic systems would be vanishingly small, thus inhibits formation of the pairing state. Thus, there must be a critical doping above which such a state is visible. 
The doping gives rise to a Fermi surface surrounding high symmetry points $\Gamma$ and $R$. When the Fermi surfaces around $\Gamma$ are with Chern number $3$ and $1$ as well as the Fermi surfaces around $R$ are with Chern numbers $2$ or $0$, the topological state with nonzero winding number emerges. 

\begin{figure}[htbp]
\centering
\includegraphics[width=0.8\textwidth]{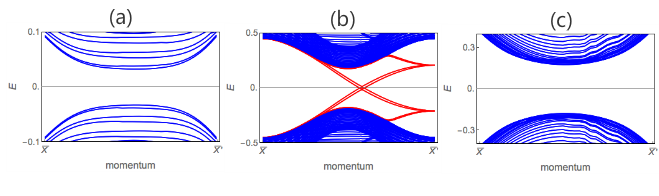}
\caption{ (Color online) The energy spectrum with open boundary condition in $y$ direction at chemical potential (a) $\mu=-1$ (b) $\mu=0.2$ (c) $\mu=1.5$. We found that edge states present at chemical potentials between $\mu=-0.5$ to $\mu=0.3$.   }
\label{RFig1}
\end{figure}

To determine the critical doping, we can either calculate the winding number or check the presence of edge state at different chemical potential.
In Fig.~\ref{RFig1}, we found that edge states present at chemical potentials between $\mu=-0.5$ to $\mu=0.3$. It indicates the critical chemical potential is $\mu=-0.5$ in our tight-binding model.

\section{construction of wavefunction at $\Gamma$ and $R$ points}\label{sec:wavefunction}
In this section, we are going to construct the wavefunctions at two high symmetry points $\Gamma$ and $R$, which are used to project the SC pairing onto the Fermi surfaces around the high symmetry points in the main text. At $\Gamma$ and $R$, the non-symmorphic symmetry can be treated as its point group component which is  independent with momentum $k$. Accordingly, we can use two independent $\bf{k}\cdot\bf{p}$ models near $\Gamma$ and $R$ instead of the continnum model.  However, the $k\cdot p$ form is usually difficult to diagonalize. According to Ref.~\onlinecite{bradlyn2016beyond}, the low energy Hamiltonian describing fourfold or sixfold degeneracies can be written in $\bf{k}\cdot\bf{S}$ form, where $\bf{S}$ is the vector of spin-3/2 or -1 matrices.  Here, we will  first project the tight-binding model into a $\bf{k}\cdot\bf{p}$ form and then project the corresponding eigenvectors onto the original $\bf{k}\cdot\bf{p}$ form. 

After choosing linear terms originating from the tight-binding Hamiltonian expanded around $\Gamma$ point ($H_\Gamma=H|_{k_x=k_y=k_z=0}+\delta k_x \frac{\partial H}{\partial k_x}|_{k_x=1,k_y=k_z=0}+\delta k_y \frac{\partial H}{\partial k_y}|_{k_x=0,k_y=1,k_z=0}+\delta k_z \frac{\partial H}{\partial k_x}|_{k_x=0,k_y=0,k_z=1}$,  we obtain a $\bf{k}\cdot\bf{p}$ Hamiltonian near $\Gamma$ as 
 \begin{align}
 \label{eq:HG}
H_\Gamma&=
   v_1(\tau_x+\tau_x\mu_x+\mu_x)\\ \nonumber
 & +v_p(\mu_y k_x+\tau_y\mu_z k_y+\tau_y\mu_x k_z)\\ \nonumber
 & +v_{r1}(\tau_y\mu_z\sigma_y+\tau_x\mu_x\sigma_z+\mu_y\sigma_x)\\ \nonumber
 &+v_{r2}(\tau_y\sigma_z+\tau_x\mu_y\sigma_x+\tau_z\mu_y\sigma_y)\\ \nonumber
&+ v_{s1}(\tau_x\sigma_xk_x+\tau_x\mu_x\sigma_y k_y+\mu_x\sigma_z k_z)\\ \nonumber
 & +v_{s2}(\tau_x\sigma_xk_x+\tau_x\sigma_y k_y+\tau_x\mu_x\sigma_x k_z)\\ \nonumber
  & +v_{s3}(\tau_z\mu_x\sigma_yk_x+\tau_x\mu_z\sigma_z k_y-\tau_y\mu_y\sigma_x k_z)\\ \nonumber
  & +3 v_2-\mu_\Gamma. \nonumber
 \end{align}
which is constraint by the little group isomorphic to point group 23.

Before deriving the eigenvector of Eq.~\eqref{eq:HG}, we construct the projection operator connect the $\bf{k}\cdot\bf{p}$ form to the $\bf{k}\cdot\bf{S}$ form with $\bf{S}$ being the vector of spin-3/2 matrices. The spin-3/2 matrices can be view as a total angular momentum as a mixture of the $p$-orbital momentum $l$=1 and the electron spin momentum 1/2. The $p$-orbital momentum can be constructed from the matrix representations of the three generators at $\Gamma$. 
In particular, the representation $G_1$ and $G_2$ of the spinless screws operator $\tilde{s}_{2x}$ and $\tilde{s}_{2z}$ are commute with the representation $G_3$ of spinless threefold rotation $\tilde{C}_{3,111}$. The commutation relations read
$G_1^2=G_2^2=G_3^3=1,\quad [G_1,G_2]=0,\quad G_1G_3=G_3G_2, \quad G_2G_3=G_3G_1G_2$. We find that these commutation relations are satisfied by the following spinless matrices 
\begin{equation}
G_1=\tau^x, \quad G_2=\mu^x\quad \text{and}\quad G_3=\left(\begin{array}{cccc}1 & 0 & 0 & 0 \\0 & 0 & 1 & 0 \\0 & 0 & 0 & 1 \\0 & 1 & 0 & 0\end{array}\right) .
\end{equation}
Here, the eigenvalues of $G_1(G_2)$ are $m_x(m_z)=\pm1$ and the corresponding eigenvectors are analogous to three $p$ orbitals and one $s$ orbitals. The $s$ orbital with $(m_x, m_z)=(-1, 1)$ corresponds to a trivial representation and the three $p$ orbitals under SU(2) invariance form a three-dimensional irreducible representation: $p_x$ orbital with $(1, -1)$, $p_y$ orbital with $(-1, -1)$, $p_z$ orbital with $(-1, 1)$ .  In particular,
 \begin{align}
 p_x=\frac{1}{2}(-1,1,1,-1) \\\nonumber
 p_y=\frac{1}{2}(-1,1,-1,1) \\\nonumber
 p_z=-\frac{1}{2}(-1,-1,1,1). \\\nonumber
 \end{align}

We express the three $p$ orbitals as linear combinations with distinct angular momentum eigenvalues ($l$, $m_l$) : $p_z$ orbital with $l=0, m_l=0$; $p_\pm=\frac{1}{\sqrt{2}}(p_x\pm i p_y)$ orbitals with $l=1, m_l=\pm 1$.  When coupled to the spin, the component of the total angular momentum along $z$ direction $m_j=\pm \frac{3}{2}, \pm \frac{1}{2}$ is good quantum number to characterize the fourfold degenerate state at $\Gamma$. And the Clebsch-Gordan coefficients are assigned as 
$p_{\frac{3}{2}}=\tilde{p}_{+}\otimes |\uparrow\rangle,
p_{\frac{1}{2}}=\sqrt{\frac{1}{3}}\tilde{p}_{+}\otimes |\downarrow\rangle+\sqrt{\frac{2}{3}}\tilde{p}_{z}\otimes |\uparrow\rangle, 
p_{-\frac{1}{2}}=\sqrt{\frac{1}{3}}\tilde{p}_{-}\otimes |\uparrow\rangle+\sqrt{\frac{2}{3}}\tilde{p}_{z}\otimes |\downarrow\rangle, 
p_{-\frac{3}{2}}=\tilde{p}_{-}\otimes |\downarrow\rangle$.

Thus we obtain the spin-3/2 matrices
\begin{align}
p_{\frac{3}{2}}&=\frac{1}{\sqrt{2}}\left\{\frac{1}{2}+\frac{i}{2},0,-\frac{1}{2}+\frac{i}{2},0
   ,\frac{1}{2}-\frac{i}{2},0,-\frac{1}{2}+\frac{i}{2},0\right\}\\\nonumber
p_{\frac{1}{2}}&=\frac{1}{\sqrt{6}}\left\{-1,\frac{1}{2}+\frac{i}{2},-1,-\frac{1}{2}+\frac{i}{2},1,\frac{1}{2}-\frac{i}{2},1,-\frac{1}{2}+\frac{i}{2}\right\}\\\nonumber
p_{-\frac{1}{2}}&=\frac{1}{\sqrt{6}}\left\{-\frac{1}{2}+\frac{i}{2},-1,\frac{1}{2}-\frac{
   i}{2},-1,-(\frac{1}{2}+\frac{i}{2}),1,\frac{1}{2}+\frac{i}{2},1\right\}\\   \nonumber
p_{-\frac{3}{2}}&=\frac{1}{\sqrt{2}}\left\{0,-(\frac{1}{2}-\frac{i}{2}),0,\frac{1}{2}-\frac{i}{2}
   ,0,-(\frac{1}{2}+\frac{i}{2}),0,\frac{1}{2}+\frac{i}{2}\right\}.
\end{align}
It is worthwhile to note that they are related to each other by $s_{2x}$ or $\mathcal{T}$.

Now we can project the eight-band $\bf{k}\cdot\bf{p}$ Hamiltonian onto this basis $(p_{\frac{3}{2}},p_{\frac{1}{2}},p_{-\frac{1}{2}},p_{-\frac{3}{2}})$. It becomes a four-dimensional $\bf{k}\cdot\bf{S}$ form. This can be understood since the Hamiltonian around $\Gamma$ point is constraint by the Little group isomorphic to point group 23. The Hamiltonian at $\Gamma$ point ($\vec{k}=0$) has four-fold degenerate bands with the same mass $m$ and it can be written into a four-dimensional linearized form around $\Gamma$.  
\begin{equation}
\label{eq:HlittleG}
H_{3/2}=m\mathbbm{1}+a\vec{k}\cdot\vec{S}+b\vec{k}'\cdot\vec{S'}+c [S_z, (S_x S_x - S_y S_y) ]
\end{equation}
where $a$, $b$ and $c$ are three independent real-valued parameters and $\mathbbm{1}$ is a four-dimensional matrices. $\vec{S}=(S_x, S_y, S_z)$ and $\vec{S}'=(S_x^3, S_y^3, S_z^3)$, in which  $S_x, S_y$, and $S_z$ are spin-$3/2$ matrices:
\begin{equation}
S_x=\frac{1}{2}\left(\begin{array}{cccc}0 & \sqrt{3} & 0 & 0 \\\sqrt{3} & 0 & 2 & 0 \\0 & 2 & 0 & \sqrt{3} \\0 & 0 & \sqrt{3} & 0\end{array}\right), \quad
S_y=\frac{1}{2i}\left(\begin{array}{cccc}0 & \sqrt{3} & 0 & 0 \\-\sqrt{3} & 0 & 2 & 0 \\0 & -2 & 0 & \sqrt{3} \\0 & 0 & -\sqrt{3} & 0\end{array}\right), \quad
S_z=\frac{1}{2}\left(\begin{array}{cccc}3 & 0 & 0 & 0 \\0 & 1 & 0 & 0 \\0 & 0 & -1 & 0 \\0 & 0 & 0 & -3\end{array}\right). \quad
\end{equation}
The two forms (Eqs.~\eqref{eq:HG} and \eqref{eq:HlittleG}) are related  with $m=-m_\Gamma + 3 v_q + v_{r1} - v_{r2} - \mu_\Gamma$, $a=\frac{1}{12} (14 v_{s1} - 9 v_{s2} + 9 v_{s3} + 4 v_\Gamma)$, $b=\frac{1}{3}(-2v_{s1}+v_{s2}-v_{s3})$ and $c=\frac{1}{6}(v_{s2}+v_{s3})$.

\begin{figure}[htbp]
\centering
\includegraphics[width=0.8\textwidth]{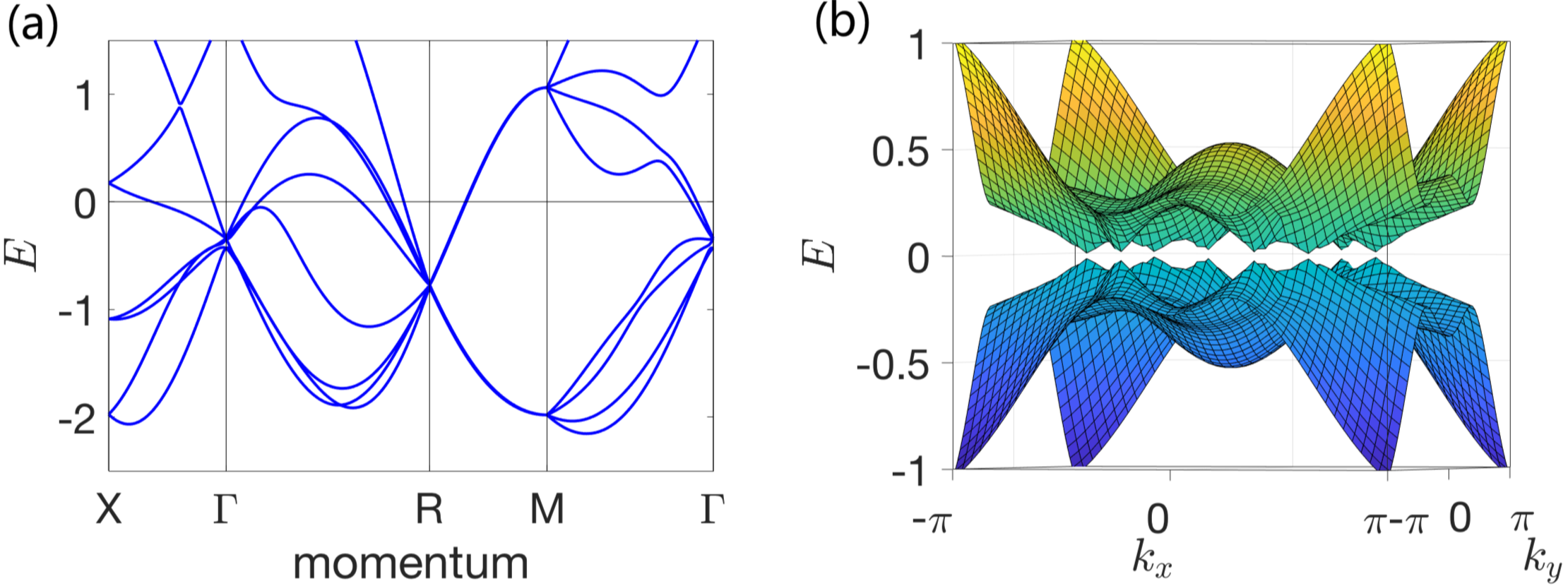}
\caption{ (Color online) (a) The bulk band structure for RhSi along high symmetry lines and (b) the surface states in the BdG spectrum with SC order parameter $\Delta=0.5$. $v_{s1}=v_{s2}=-v_{s3}=-0.76$ and the other parameters are the same with Fig.~\ref{Fig2}(b).}
\label{SFig2}
\end{figure}

It is easy to find the eigenstates at $b=c=0$ and $m=0$, by setting  $v_{s1}=v_{s2}=-v_{s3}$  and $\mu_\Gamma=-m_\Gamma + 3 v_q + v_{r1} - v_{r2}$. 
Note that although the parameters are changed to $v_{s1}=v_{s2}=-v_{s3}=-v_p$ , the energy degeneracy at $\Gamma$ and $R$ are kept, as shown in Fig.\ref{SFig2}(a).  And the existence of TSC for the new set of parameters is established by Fig.~\ref{SFig2}(b).

With $v_{s1}=v_{s2}=-v_{s3}$  and $\mu_\Gamma=-m_\Gamma + 3 v_q + v_{r1} - v_{r2}$, the two eigenvectors of Eq.\eqref{eq:HlittleG} with positive eigenvalues $\frac{3}{2}$ and $\frac{1}{2}$ can be obtained
\begin{align}
\phi_{\Gamma,3/2}=\frac{e^{-3 i \phi }}{2\sqrt{2}}\bigg\{&-\left(\cos\frac{\theta }{2}-\sin\frac{\theta
   }{2}\right)^3,-\sqrt{3} e^{i \phi } (\sin\theta-1) \left(\sin
 \frac{\theta }{2}+\cos\frac{\theta }{2}\right),\\ \nonumber
 &\sqrt{3} e^{2i
   \phi }\frac{ (\sin\theta-1)}{ \cos\theta } \left(\sin\frac{\theta }{2}+\cos
\frac{\theta }{2}\right)^3,e^{3i
   \phi }\left(\sin\frac{\theta }{2}+\cos
   \frac{\theta }{2}\right)^3\bigg\} \\\nonumber
   \phi_{\Gamma,1/2}=\frac{e^{-3 i \phi }}{4}\bigg\{&- \sqrt{3} (\sin\theta-1) \left(\sin\frac{\theta
   }{2}+\cos\frac{\theta }{2}\right),- e^{ i \phi } (3 \sin
   \theta+1)\left (\cos\frac{\theta }{2}-\sin\frac{\theta
   }{2}\right),\\\nonumber
   & e^{2i \phi } (3 \sin\theta-1)\left (\sin
\frac{\theta }{2}+\cos\frac{\theta }{2}\right),e^{3i \phi }\frac{\sqrt{3}
   \cos ^2\theta }{\cos \frac{\theta }{2}- \sin\frac{\theta
   }{2}}\bigg\}.
\end{align}
 Accordingly, the eigenstate of Eq.~\eqref{eq:HG}  can be easily obtained $|\psi_{\Gamma,i} \rangle=|\phi_{\Gamma,i} \rangle\langle p_{\frac{3}{2}}, p_{\frac{1}{2}}, p_{-\frac{1}{2}}, p_{-\frac{3}{2}}|$. And $(|\psi_{\Gamma,3/2}\rangle$ and $|\psi_{\Gamma,1/2}\rangle)$ are the time-reversal partners of each other after fixing the phase. 

Similarly, we can expand the Hamiltonian near $R$ point and obtain
  \begin{align}
 \label{eq:HR}
H_R&=
+v_p(\tau_y\mu_z k_x+\tau_y\mu_x k_y+\mu_y k_z)\\ \nonumber
 & +v_{r3}(\tau_y\mu_z\sigma_x+\tau_y\mu_x\sigma_y+\mu_y\sigma_z)\\ \nonumber
&+ v_{s1}(\mu_x\sigma_zk_x+\tau_x\mu_x\sigma_x k_y+\tau_y\mu_x\sigma_y k_z)\\ \nonumber
 & +v_{s2}(\tau_x\sigma_yk_x+\tau_x\mu_x\sigma_z k_y+\mu_x\sigma_x k_z)\\ \nonumber
  & +v_{s3}(\tau_x\mu_z\sigma_zk_x-\tau_y\mu_y\sigma_x k_y+\tau_z\mu_x\sigma_y k_z)\\ \nonumber
  & -3 v_2-\mu_R. \\ \nonumber
 \end{align}
  
  At $R$,  the Hamiltonian is six-fold degenerate and a three-dimensional irreducible representation can be given by  the three generators $
  s_{2xR}=\{C_{2x}|\frac{1}{2}\frac{3}{2}0\}, s_{2yR}=\{C_{2y}|0\frac{1}{2}\frac{3}{2}\}, \quad\text{and}\quad s_{3R}=\{C^{-1}_{3,111}|010\}$. The representation of generators $G_1$, $G_2$ and $G_3$ satisfy $G_3^3=G_1^2=G_2^2=1, G_1G_2=G_2G_1, G^{-1}_3G_1G_3=G_2, G^{-1}_3G_2G_3=G_1G_2$,  and thus are different from the representation of generators at $\Gamma$.

We take one of the $p's$ in the three-dimensional irreducible basis $p_R$ and its $C_3$ rotation partners $|p_R\rangle, G_3 |p_R\rangle\quad \text{and} \quad  G_3 ^2|p_R\rangle$
form a 3d representation, while their time-reversal partners ($\mathcal{T}|p_R\rangle$, $G_3\mathcal{T}|p_R\rangle$ and  $G_3 ^2\mathcal{T}|p_R\rangle$) consist another 3d representation. In combination, a six-dimensional (6D) basis is formed
  \begin{equation}
  |P\rangle=\{|p_R\rangle, G_3 |p_R\rangle,  G_3 ^2|p_R \rangle, \mathcal{T}|p_R\rangle, G_3 \mathcal{T}|p_R\rangle, G_3 ^2\mathcal{T}|p_R \rangle\}.
  \end{equation}
which gives a 6D representation of the Hamiltonian in a $\delta\bf{k}\cdot\bf{S}$ form.

To construct this 6D representation for space group 198 at $R$, we note that the unitary subgroup of the little group is isomorphic to the little group of space group 199 and satisfies
\begin{equation}
\label{eq:HlittR}
H_{198}({\bf k})=\begin{pmatrix} H_{199}(a,{\bf k}) & bH_{199}(1,{\bf k})\\ b^*H_{199}(1,{\bf k}) & -H_{199}^*(a,{\bf k})\end{pmatrix}
\end{equation}
where the Hamiltonian for SG 199 is given by Gell-Mann matrices as
 \begin{equation}
 H_{199}=\begin{pmatrix}0 & a k_x & a^* k_y \\
 a^* k_x &0 & a k_z\\
 a k_y & a^* k_z &0 
 \end{pmatrix}
 \end{equation}
 with $a=|a|e^{i\eta}$ is a complex parameters.

When  $\eta=\frac{\pi}{6}$,  we note that  Eq.\eqref{eq:HlittR} can be exactly solvable. The eigenvector below the fermi surfaces  is
\begin{align}
|\phi_{R, -2}\rangle=&\bigg\{0,0,0,\frac{(1-i) \left(3+\sqrt{3}\right) \cos \theta }{\left(3+(2+i) \sqrt{3}\right) (\sin \theta  \cos (\phi
   )+i \sin (\phi ))},\frac{\left(\sqrt{3}-i\right) (\cos (\phi )+i \sin \theta  \sin (\phi ))}{2 \sin \theta  \cos
   (\phi )+2 i \sin (\phi )},1\bigg\}
\end{align} 

When the eigenvectors projected onto $p$ orbitals basis, they are written into eight-dimensional vectors $|\psi_{R, 1} \rangle=|\phi_{R,-2} \rangle \langle P|$.
 While $|\psi_{R,-1}\rangle$ can be utilized to project the Hamiltonian to one of the Fermi surfaces with $\mathcal{C}=-2$ around $R$, its time-reversal partner $\mathcal{T}|\psi_{R,-1}\rangle$ project to another Fermi surface.

%\bibliography{weyl}
\end{document}